
\input phyzzx.tex
\input phyzzx.plus
\def\a{\alpha}
\def\b{\beta}

\def\d{\delta}

\def\z{\zeta}

\def\r{\rho}

\def\k{\kappa}
\def\l{\lambda}

\def\O{\Omega}
\def\s{\sigma}

\def\pa{\partial}
\def\na{\nabla}
\def\hg{\hat g}

\def\ov{\overline}

\def\zp{\zeta_+}
\def\zm{\zeta_-}

%
\def\ap#1{{\it Ann. Phys.} {\bf #1}}
\def\cmp#1{{\it Comm. Math. Phys.} {\bf #1}}
\def\cqg#1{{\it Class. Quantum Grav.} {\bf #1}}
\def\pl#1{{\it Phys. Lett.} {\bf B#1}}

\def\prd#1{{\it Phys. Rev.} {\bf D#1}}

\def\np#1{{\it Nucl. Phys.} {\bf B#1}}

 \REF\kk{ K.V. Kucha\v{r},
``Time and Interpretations of Quantum Gravity'',
in {\it Winnipeg 1991, Proceedings, General relativity and
relativistic
astrophysics}; \hfill\break
C.J. Isham, ``Canonical Quantum Gravity and the Problem of Time'',
Presented at 19th International Colloquium on Group Theoretical
Methods in
Physics, Salamanca, Spain, 29 Jun - 5 Jul 1992, gr-qc/9210011.}

\REF\hawk{S. W. Hawking, \cmp{43},  199 (1975).}

\REF\tHooft{G. 't Hooft, \np{335}, 138 (1990), and ``Unitarity of the
Black
Hole S-Matrix'', THU-93/04.}

\REF\tj{T. Jacobson, \prd{48}, 728 (1993).}

\REF\stu{L. Susskind, L. Thorlacius, J. Uglum, ``The Stretched
Horizon and
Black Hole Complementarity'', SU-ITP-93-15, hep-th/9306069.}

\REF\banks{V. Lapchinsky and V. Rubakov, {\it Acta Phys. Pol.} {\bf B
10}, 1041
 (1979);\hfill\break T. Banks, \np{249}, 332 (1985);\hfill\break P.D.
D'Eath
and J. Halliwell, \prd{35}, 1100 (1987);\hfill\break
R. Brout, {\it Found. Phys.} {\bf 17}, 603 (1987);\hfill\break
R. Brout and G. Venturi, \prd{39}, 2436 (1989);\hfill\break
T.P. Singh and T. Padmanabhan, \ap{196}, 296 (1989);\hfill\break
C. Kiefer, ``The Semi-classical Aproximation to Quantum Gravity'', to
appear in
{\em Canonical Gravity- from Classical to Quantum}, ed. by J. Ehlers
and H.
Friedrich (Springer Verlag, 1994), FREIBURG-THEP-93-27,
gr-qc/9312015.}

\REF\call{C.G. Callan, S.B. Giddings, J.A. Harvey, and A. Strominger,
\prd{45}, R1005 (1992).}

\REF\sda{S. P. de Alwis, \pl{289}, 278 (1992); \pl{300}, 330 (1993);
\prd{46},
5429 (1992).}

\REF\bc{A. Bilal and C. Callan, \np{394}, 73 (1993).}

\REF\shda{S. P. de Alwis, \pl{317}, 46 (1993).}

\REF\bd{N. D. Birrel and P. C. W. Davies, {\it Quantum Fields in
Curved Space}
(Cambridge University Press, Cambridge, 1982).}

\REF\rst{J. Russo, L. Susskind, L. Thorlacius, \pl{292}, 13 (1992);
 L. Susskind, L. Thorlacius, \np{382}, 123 (1992);  L. Susskind, L.
Thorlacius,
\prd{46}, 3444 (1992).}

\REF\holland{Peter R. Holland, {\it  The Quantum Theory of Motion}
(Cambridge
University Press, Cambridge, 1993).}

\REF\sh{A. Strominger, \prd{48}, 5769(1993);\hfill\break S. W.
Hawking, ``The
Superscattering Matrix for Two-dimensional Black Holes'', DAMTP
preprint,
hep-th-9401109.}

\REF\dea{S. P. de Alwis, \prd{49}, 941 (1994).}

\REF\rt{T. Regge and C. Teitelboim, \ap {88},  286 (1974).}

\REF\jb{J. S. Bell,  {\it Speakable and Unspeakable in Quantum
Mechanics}
(Cambridge University Press, Cambridge, 1987).}

\REF\gh{M. Gell-Mann and J. B. Hartle, in {\it Complexity, Entropy,
and
Information (Vol. VIII in the SFI Studies in the Science of
Complexity)}, ed.,
W. Zurek (Addison Wesley, Redwwod City, CA, 1990).}

\REF\vink{J. Kowalski-Glikman and J. Vink, \cqg{7}, 901
(1990);\hfill\break
J. Vink, \np{369}, 707 (1992).}

\REF\hh{J. Hartle and S. Hawking, \prd{28}, 2960 (1983),}

\REF\vil{A. Vilenkin, \prd{37}, 888 (1988).}

\REF\hori{T. Hori, ``Exact Solutions to the Wheeler-deWitt Equation
of
Two-Dimensional Dilaton Gravity'',  TEP-11,
hep-th/9303049;\hfill\break
D. Louis-Martinez, J. Gegenberg, G. Kunstatter, \pl{321}, 193
(1994).}

\REF\wit{E. Witten, \cmp{117} 353 (1988).}

\REF\reviews{J.A. Harvey and A. Strominger, ``Quantum Aspects of
Black Holes''
in {\it Recent Directions in Particle Theory},  J. Harvey and J.
Polchinski,
eds.  (World Scientific, Singapore, 1993);\hfill\break
Steven B. Giddings, ``Toy Models for Black Hole Evaporation'',
UCSBTH-92-36,
Presented at International Workshop on Theoretical Physics: 6th
Session:
String Quantum Gravity and Physics at the Planck Energy Scale, Erice,
Italy.}

\pubnum {COLO-HEP-333\cr}
\date={March 1994}
\titlepage
\vglue .2in
\centerline{\bf  The Problem of Time and Quantum Black Holes}
\author{S. P. de Alwis and D. A. MacIntire }
\address{Dept. of Physics, Box 390,\break
University of Colorado,\break Boulder, CO
80309}
\vglue .2in
\centerline{\caps ABSTRACT}
We discuss the derivation of the so-called semi-classical equations
for both
mini-superspace and dilaton gravity. We find that there is no
systematic
derivation of a semi-classical theory in which quantum mechanics is
formulated
in a space-time that is a solution of  Einstein's equation, with the
expectation value of  the matter stress tensor on the right-hand
side. The
issues involved are related to the well-known problems associated
with the
interpretation of  the  Wheeler-deWitt equation  in quantum gravity,
including
the problem of time. We explore the de Broglie-Bohm interpretation of
quantum
mechanics (and field theory) as a way of spontaneously breaking
general
covariance, and thereby giving meaning to the equations that many
authors have
been using to analyze black hole evaporation. We comment on the
implications
for the ``information loss" problem.

\endpage
\chapter{\bf Introduction}

It is usually assumed that the problem of time in quantum gravity is
irrelevant
 to the questions associated with black hole radiation. In particular
it is
thought that an S-matrix (or super scattering operator \$) can be
defined and
that the whole issue can be discussed within the orthodox
interpretation of
quantum mechanics (with classical apparatus and a classical
observer).   The
implicit assumption seems to be that the analyses need not in
principle be
different from that of  some particle physics scattering experiment.

However  a notion of time evolution  is essential to the definition
of a
scattering (or super-scattering) operator, and the  question of
whether
information is lost or not is well posed only within that context.
It is also
well known that in the usual formulation of quantum gravity  (with
Dirac or
BRST quantization) there is no  time evolution. As discussed in
detail in the
reviews [\kk] there are serious problems with all attempts to resolve
this
question. Nevertheless, it is often asserted that as far as the black
hole
problem is concerned this is not an issue.  The reasons are not often
explicitly stated, but  one or other of the  following assumptions
are implicit
in most analyses of the problem:

a) Hawking's original calculation [\hawk ] is semi-classical and one
should be
able to derive a notion of time in this approximation.

b) The black hole space-time is open and asymptotically flat. In this
case
there is a non-vanishing Hamiltonian associated with spatial
infinity. This
should enable one to
define a notion of  time evolution and an S (or \$)-matrix.

Now if the semi-classical theory is  internally self consistent one
may ignore
the problem of time and leave the issue of how to derive the theory
from
quantum gravity  to future work. However the semi-classical theory is
non-linear (since it involves the expectation value of the matter
stress
tensor) and thus violates the superposition principle. Thus it is
unclear how
one could use standard S-matrix ideas.  Furthermore, it has become
increasingly
clear that the issue of information loss cannot be resolved one way
or the
other without some understanding of what happens at short distance
scales (such
as distances within a Planck length of the horizon)\foot{See for
example
[\tHooft, \tj, \stu].}. This is clearly the regime of  quantum
gravity. Since
there is no well defined theory at these scales (in four
dimensions)\foot{String theory may be the answer, but at present
there is no
real understanding of anything beyond perturbation theory around flat
backgrounds.}, one can  make assumptions about this regime in
accordance with
ones prejudices.  However, there is still a serious problem of
principle
involved here.  Whatever the final version of quantum gravity is
like, as long
as  the requirement of general covariance is imposed, the quantum
theory would
seem to have the problem of time. The question that arises then is
whether the
semi-classical  analyses can emerge from the quantum theory in some
limit, at
least for wave functions (such as that of a coherent state) that are
amenable
to a classical interpretation.

There are arguments in the literature that attempt to derive
semi-classical
physics, and in  particular the (functional) Schr\"{o}dinger
equation, from the
Wheeler-deWitt  (WdW) equation [\banks]. These  are sometimes used to
justify
the statement that the semi-classical treatment of the black hole
problem can
be derived, in the large Planck mass limit, from quantum gravity. A
careful
examination of these arguments however reveals that there is no {\it
systematic} way in which the equations of semi-classical black hole
physics can
be derived. In fact the usual argument does give a Schr\"{o}dinger
equation,
but in a background which is a solution of the vacuum Einstein
equations, i.e.
it is  independent of the (quantum) state of matter. Both in the
black hole
case and in cosmological applications, however, the geometry is
supposed to be
determined by (the expectation value of) the matter stress tensor, so
the usual
argument is clearly inadequate. Since the original work on this
question there
have been attempts to remedy this problem by various means.  In the
first part
of this paper we will argue that none of these are satisfactory.

We will first discuss the issue in mini-superspace.  The conceptual
problem
that we face is already manifest at this level. However, the recent
development
 of two-dimensional dilaton-gravity field theories which exhibit
classical
dynamical black hole solutions (Callen, Giddings, Harvey, and
Strominger
(CGHS), reference [\call ]) gives us a much more interesting toy
model within
which these questions can be investigated. In fact the work of
references
[\sda, \bc]  and [\shda ] established that there is a class of
quantum CGHS
theories which are exactly solvable (quantum) conformal field
theories (CFT) in
two dimensions. These theories are well defined quantum gravity
theories which
have non-trivial (dynamical black hole) solutions in the classical
limit.
Thus these models are ideal laboratories for the study of the passage
from the
exact quantum to the semi-classical physics of black holes. Some
comments on
the issues involved have already been published in [\shda ]. In that
paper it
was pointed out that the usual (exact) quantization of the theory
(which in
this case is a  2d conformal field theory) completely obscures the
semi-classical picture.  In this paper we elaborate further on this
and in
particular address two new questions. The first is whether there is
an
alternative to Dirac quantization (such as light cone gauge) which
can solve
the problem.   Our conclusion is that it cannot. The second is
whether the fact
that we have an open system can resolve the problem as in (b) above.
Here too
the conclusion is negative. The boundary Hamiltonian is irrelevant
for the
local dynamics of the quantum field theory and hence has no effect at
all on
the problem of time.

A possible resolution of these issues is to give up the superposition
principle
for quantum
gravity. In fact, as is well known (see for example [\kk]), in the
attempts to
derive the semi-classical equations from the Wheeler-deWitt equation
this
principle is effectively abandoned. The argument may be made that the
wave
function of the universe is unique and one does not have the usual
reasons for
superposing states, which are valid only for  the quantum mechanics
of
subsystems. Indeed the wave function of the universe as it evolves
would give
rise to the experimental situations which behave like superpositions
of
different states of subsystems. At the fundamental level there may be
no need
to satisfy the superposition principle. If one accepts this argument
then one
loses the need to impose the WdW equation (as a linear constraint).
Of course
it is still necessary to recover Einstein's equations in the
classical limit.
This may be done by imposing the constraint as an expectation value.
In other
words, the constraint operators do not annihilate the physical
states. Instead,
the only permissible states (i.e. states with the correct classical
limit,
giving only geometries allowed by Einstein's theory) are those in
which the
constraint operators have expectation value zero.  Since this is a
non-linear
condition, a linear superposition of such states is in general not a
physical
state.

Such an ansatz would constitute a spontaneous breakdown of general
covariance.
In [\shda]
it was argued that this  may be what one needs in order to have the
standard
semi-classical picture. The semi-classical equations that have been
used
hitherto in all the interesting applications of quantum gravity (e.g.
see
[\bd]) are then completely justified, if  at the quantum level one
takes a
coherent state. In the case of the solvable quantum gravity theory
[\sda, \bc,
\shda ] (defined in the above fashion) that emerges from  the
classical CGHS
model, we see in fact that the semi-classical analysis is exact.
However this
is true only for those theories in which there is no boundary in
field space
(see the third paper of [\sda] and [\shda]). In theories with a
boundary [\rst]
(if they are well defined!) this is not so, but the semi-classical
equations
will emerge in the large $N$ limit in just the way that one expects
in ordinary
(non-generally covariant) quantum mechanics. This is to be contrasted
with the
fact that if the usual  linear constraint equations are used, then
there is no
way of recovering the semi-classical limit in the large $N$ limit.

This derivation of  semi-classical  physics involves abandoning the
linearity
of quantum mechanics at the fundamental level. While this may be
justified in
so far as one takes the point of view that there is a unique wave
function of
the universe, we believe that it  is worthwhile exploring an
alternative. To
this end we study the de Broglie Bohm interpretation of quantum
mechanics
\foot{For a comprehensive review see reference [\holland].}. In this
interpretation there are definite orbits for the  fields of the
theory which
are guided by the quantum  wave functional. The latter also gives the
possible
distribution of the initial values of such fields. In the context of
quantum
gravity, picking an orbit  means picking a particular space-time and
is
tantamount to spontaneously breaking  general covariance.  The
equations for
these orbits are in fact the semi-classical equations corrected by
certain
higher order terms. We believe that this interpretation gives a very
natural
way of understanding the emergence of  semi-classical physics.
However it must
be stressed that what one gets is not what is usually used in quantum
gravity
(which is in fact more properly called the Born-Oppenheimer
approximation), but
the standard $O(\hbar )$ expansion. In other words, one has (semi-)
classical
equations explicitly modified  by quantum corrections.

In the next section of this paper we discuss the arguments for the
emergence of
semi-classical physics from the WdW equations, and conclude that
there is no
systematic derivation of the former. In the third section we discuss
this issue
in the CGHS theory and come to the same conclusion. In the fourth
section we
show that both reduced phase space quantization, and the existence of
a
boundary Hamiltonian, are irrelevant to the questions that we are
addressing.
In the fifth section we present a possible resolution of these
issues, and in
the concluding section we summarize our arguments.

\chapter{The Problem of Time and the Semi-classical Approximation in
Mini-superspace}

In this section we will review the derivation the semi-classical
limit of the
Wheeler-deWitt equation for the example of
mini-superspace\refmark{\banks}.  We
consider a homogenous, isotropic space-time with the metric$$ds^2 =
-N(t)^2
dt^2 + a(t)^2 d\Omega_3^2$$where the lapse function $N$ and the scale
factor
$a$ only depend upon $t$, and $d\O_3^2$ is a metric on three space.
The action
for gravity coupled to a homogeneous scalar field $\phi(t)$ is
$$S=\int dt \left[-a^3  M^2 \left\{{1\over 2N} \left( {\dot a\over
a}\right)^2
+ N\, V_G(a)\right\}+ a^3 \left\{{\dot \phi^2\over 2 N} - N\,
V_m(\phi)\right\}\right]$$where
$$V_G(a) = -{1\over 2} k a^{-2} + {1\over 6} \Lambda,$$ and $M$ is
the Planck
mass.  The equations of motion from this  action (in the $N=1$ gauge)
are:
$$\eqalignno
{\delta N : \qquad & -M^2 \left({a \dot a^2\over 2} - a^3
V_G(a)\right) +
a^3\left({\dot \phi^2\over 2} + V_m(\phi)\right) = 0&\eqnalign{\eqf}
\cr
 \delta a  : \qquad & M^2\left\{{d\over dt} (a \dot a) - {\dot
a^2\over 2} -
\partial_a \left(a^3 V_G(a)\right) \right\} + 3
a^2\left({\dot\phi^2\over 2} -
V_m(\phi)\right) =0&\eqnalign{\eqs}\cr
\delta \phi : \qquad & {d\over dt}\left(a^3 \dot\phi\right) + a^3
\partial_\phi
V_m(\phi) = 0.&\eqnalign{\eqt}\cr}$$The first equation is the
constraint that
the total energy of the system is zero.

In terms of the Hamiltonian for this system
$$H = -{1\over 2 a M^2}  p_a^2 +  a^3 M^2 V_G(a)   + {1\over 2 a^3}
p_\phi^2 +
a^3 V_m(\phi)\eqn\Ham$$with $$p_a=-M^2 a \dot a\qquad p_\phi = a^3
\dot \phi ,
\eqn\CanP$$equation \eqf\ is the (secondary) constraint $$H \approx
0.$$

Upon Dirac quantization of this system one has the physical state
condition on
the `wave function for the universe':
$$\hat H \ket{\Psi} =0.\eqn\WdW$$This constraint\foot{Equation \WdW\
is the
statement that the wave function $\Psi$ is invariant under time
reparametrizations.  In the full theory the constraints are a
reflection of
general covariance at the quantum level.} implies$$i \hbar
{\partial\over
\partial t} \ket{\Psi} = \hat H\ket{\Psi} = 0.\eqn\Sch$$That is, the
wave-function of the universe is a stationary state with zero energy,
and there
is no Schr\"odinger evolution of the physical states.  It also
implies that any
dynamical variable  must commute with the Hamiltonian, on the space
of physical
states, and thus also be time independent.

Is it possible to formulate a theory with Schr\"odinger  time
evolution in the
approximation that $M\gg \vev{H_m}$, where $H_m$ is the matter
Hamiltonian?  In
the Schr\"odinger representation the constraint equation is the WdW
equation,
$$\hat H \Psi (a,\phi) =\left\{ {\hbar^2\over 2 M^2} {1\over a^2}
\partial_a (
a \partial_a)
- {\hbar^2\over 2 a^3} \partial_\phi^2 + a^3 \left( M^2 V_G +
V_m\right)\right\} \Psi (a,\phi)=0.$$ In the above we have resolved
the
ordering ambiguity in the $p_a$ term by requiring that the
Hamiltonian be
hermitian in the inner product $$\braket{\Psi}{\Phi}=\int da\,
d\phi\, a^2
\Psi(a,\phi)^* \Phi(a,\phi).$$We now write $$\Psi(a,\phi) = R(a,\phi)
e^{i
S(a,\phi)/\hbar}$$where $R$ and $S$ are real functions.  Substituting
in the
WdW equation we find equations for the real and imaginary parts:
$$-{1\over 2 a M^2} (\partial_a S)^2 + {1\over 2 a^3} (\partial_\phi
S)^2 +
a^3\left[M^2 V_G + V_m\right] + { \hbar^2\over 2 a^3 R}\left[{1\over
M^2}
(a\partial_a)^2 R - \partial_\phi^2 R\right] = 0, \eqn\RePart$$
$$  \partial_\phi (R^2 \partial_\phi S) - {1\over M^2} a \partial_a
(R^2 a
\partial_a S) =0.\eqn\ImPart$$

Expanding in powers of $M$,
$$\eqalign{S(a,\phi) &= M^2 S_{-1}(a,\phi) + S_0(a,\phi) + M^{-2}
S_1(a,\phi)
\ldots\cr
R(a,\phi) &= R_0(a,\phi) + M^{-2} R_1(a,\phi) + \ldots, \cr}$$we find
the
equations
$$\eqalignno{{\cal O}(M^4):&\qquad 0 = \partial_\phi S_{-1}
&\eqnalign{\OMfour}\cr
{\cal O}(M^2):&\qquad 0 =  -{1\over 2 a} (\partial_a S_{-1})^2 + a^3
V_G.
&\eqnalign{\OMtwo}\cr}$$

Defining $$R_0(a,\phi) = {1\over \sqrt{a \partial_a S_{-1}(a)}}\,
r_0(a,\phi)$$
and $$\chi (a,\phi) = r_0(a,\phi) e^{i S_0(a,\phi)/\hbar}$$we have
$$\Psi(a,\phi) = {e^{i M^2 S_{-1} (a)/\hbar}\over \sqrt{a \partial_a
S_{-1}(a)}}\,  \left[\chi (a,\phi)+O(M^{-2})\right],$$where from
equation
\OMfour\ $S_{-1}$ only depends upon $a$.  The ${\cal O}(M^0)$
equation can now
be written as
$$i  \hbar {\partial \chi (a,\phi) \over \partial T} = H_m
\chi(a,\phi)
\eqn\WKBSch$$
where
$${\partial\over \partial T } \equiv -{1 \over a} (\partial_a S_{-1}
){\partial
\over \partial a}\eqn\Time $$ and
 $$ H_m \equiv -{\hbar^2\over 2 a^3} \partial_\phi^2 + a^3
V_m.\eqn\matter$$
{}From this we have
$${\partial a\over \partial T} =\dot a = -{1\over a} \partial_a
S_{-1}(a),\eqn\aorbit$$
thus defining a definite orbit for the scale factor.

We have therefore succeeded in obtaining a Schr\"odinger  equation
for the
evolution of the matter wave function.  However, the Hamilton-Jacobi
equation
\OMtwo\ which determines the metric (\ie \ $a$) is equivalent to the
{\em
vacuum Einstein equation}.   The usual semi-classical equation
$$G_{\mu\nu} =
8\pi G \vev{T_{\mu\nu}}$$ would have, instead of equation \OMtwo,
resulted in
an equation of the form
$$-{1\over 2 a} (\partial_a S_{-1})^2 + a^3 V_G = -{1\over M^2}
\bra{\chi}H_m\ket{\chi}.\eqn\SCone$$
What one needs in order to justify the usual semi-classical picture
is  a
derivation of the Schr\"odinger equation in a background determined
by the
above equation.  In order to see whether this is possible we will use
\SCone\
in the WdW equation, and  determine the conditions under which the
resulting
equation for the matter wave function is consistent with the
Schr\"odinger
equation.

Let us therefore write $\Psi$ as
$$\Psi (a,\phi) = e^{{1\over \hbar} M^2 S_{-1}(a)} \tilde
\chi(a,\phi)=
e^{{1\over \hbar} M^2 S_{-1}(a)} f(a) \chi (a,\phi)$$
where $S_{-1}(a)$ is defined to be a $\phi$ independent
($\partial_\phi S_{-1}
=0$) solution of \SCone.  Substituting the above expression in the
WdW
equation gives
$${i \hbar\over a} \partial_a S_{-1} \, \partial_a \tilde \chi  -
{M^2 \over 2
a}\left(\partial_a S_{-1}\right )^2 \tilde \chi+ a^3 M^2 V_G
\,\tilde\chi + {i
\hbar \over 2 a^3} \tilde\chi \,(a\partial_a)^2 S_{-1} +
{\hbar^2\over 2 M^2
a^3} (a\partial_a)^2 \tilde\chi  + H_m\tilde\chi =0.$$Using \SCone\
and
dividing by $f$ we obtain

\vbox{
$$\left[H_m - \vev{H_m}\right]\chi + i\hbar\left[{1\over
a^3}\partial_\a S_{-1}
-{i\hbar\over M^2 a^3} {\partial_\a f\over f}\right] \partial_\a \chi
+
{\hbar^2 \over 2 M^2 a^3} \partial^2_\a \chi $$
$$+\  i\hbar \left[{1\over a^3} \partial_\a S_{-1} {\partial_\a f
\over f} +
{1\over 2 a^3} \partial^2_\a S_{-1} - {i\hbar \over 2 M a^3}
{\partial^2_\a
f\over f} \right] \chi  =0 \eqn\WdWone $$
}
with $$\partial_\a \equiv a\partial_a.$$

The function $f(a)$ will be defined by requiring that the last term
in brackets
is zero:
$${1\over a^3} \partial_\a S_{-1} {\partial_\a f \over f} + {1\over 2
a^3}
\partial^2_\a S_{-1} - {i\hbar \over 2 M a^3} {\partial^2_\a f\over
f}=0.$$One
now defines the time derivative as before$$\partial_T \equiv -{1\over
a}
\partial_a S_{-1} \partial_a = -{1\over a^3} \partial_\a S_{-1}
\partial_\a
$$so that \WdWone\ becomes
$$\left[H_m - \vev{H_m}\right]\chi  -i\hbar {\partial
\chi\over\partial T} +
{\hbar^2 \over 2 M^2 a^3} \partial^2_\a \chi + {\hbar^2\over M^2 a^3}
{\partial_\a f\over f} = 0.\eqn\WdWtwo$$We now assume the
Schr\"odinger-like
equation
$$\left[H_m - \vev{H_m}\right]\chi = i\hbar {\partial
\chi\over\partial
T},\eqn\Sch$$and ask whether the remaining terms in equation \WdWtwo\
are small
compared with the terms in this equation\foot{Note that we can
redefine $\chi
=\chi_s \exp{\left\{ {1\over \hbar} \int^t \vev{H_m} dt \right\}}$
and obtain
the Schr\"odinger  equation for $\chi_s$.  In this case the
non-linear
character of the equation is hidden in the phase redefinition (see
Brout and
Venturi in reference [\banks]).}.

For simplicity let us assume that $V_G=0$ (e.g. late universe with
$\Lambda=0$).  This implies from eq. \SCone,
$$\partial_\a S_{-1} = {a^{3/2} \sqrt{2 \vev{H_m}} \over M}.$$If we
also assume
that $a$ is large, we have
$$\partial_\a S_{-1} = {a^{3/2} \sqrt{2 \vev{H_m}} \over
M}\longrightarrow
{a^{3} \sqrt{2 \vev{V_m}} \over M}.$$
After some calculation  one finds that the Schr\"odinger-like
equation \Sch\ is
consistent in the above sense, provided $$M\sqrt{\vev{H_m}} a^{3/2}
\quad
\mathrel{\mathop{\longrightarrow}_{\scriptstyle a\rightarrow \infty
}} \quad
M\sqrt{\vev{V_m}} a^3 \gg 1.$$

There are several discussions in the literature that are similar to
the above.
However we feel that it was necessary to go over this in some detail,
to
demonstrate that the derivation of semi-classical physics is far from
clear
cut. It is not by any means a systematic approximation from the exact
theory
even in this highly simplified model.

\chapter{Wheeler-deWitt Equation  in 2-d Quantum Dilaton Gravity}

The recent revival of interest in the problem of quantum radiation
from black
holes is almost entirely due to the construction of a two-dimensional
model of
dilaton gravity coupled to matter by Callan, Giddings, Harvey, and
Strominger
[\call]. The hope was that in a context where the  the intractable
ultra-violet problems of four-dimensional gravity were absent, the
conceptual
issues associated with Hawking radiation could be addressed and
resolved.
Notwithstanding claims to the contrary (eg, Strominger, Hawking [\sh]
) we
believe that even within this highly simplified context the problem
has not
been resolved. Our aim here is to demonstrate clearly that this is so
precisely
because the question is not well posed.  Indeed we believe that the
role of the
CGHS model is that it gives a well defined context within which we
can show
precisely what   problems arise in formulating the question.

The model is defined by the action,
$$S={1\over 4\pi}\int d^2\s\sqrt{-g}\left[e^{-2\phi}(R+4(\na\phi
)^2+4\l^2)-{1\over
2}\sum_{i=1}^N (\na f_i)^2 \right].\eqn\cghs$$
In the above, in addition to the metric $g$ and dilaton $\phi$
fields, there
are $N$ conformal scalar fields $f_i$. The quantum theory may be
defined by the
path integral after gauge fixing to the conformal gauge
$g_{\a\b}=e^{2\rho}\hg_{\a\b}$, where $\hg$ is a fiducial metric with
unit
determinant (say). The independence of the functional integral from
the
fiducial metric then leads to the requirement that the conformal
gauge fixed
theory with the translationally invariant measures be a conformal
field theory
(CFT) [\sda]. This may then be written in terms of the Liouville-like
action
[\sda,\bc],
$$S={1\over 4\pi}\int d^2\s \left[\mp\pa_{+}X\pa_{-}
X\pm\pa_{+}Y\pa_{-}Y+\sum_{i=1}^N
\pa_{+}f_i\pa_{-}f_i+2\l^2e^{\mp\sqrt{2\over
|\k|}(X\mp Y)}\right]. \eqn\action$$Here the upper and lower signs
correspond
to $\k<0$ and $\k>0$, respectively, and $\k ={N-24\over 6}$, $N$
being the
number of matter fields\foot{ Note that this definition has the
opposite sign
to the $\k $ defined in [\sda].   Also, henceforth we will assume
$\k>0$.}.
The field variables in the above are related to the original
variables $\phi$
and $\rho$ that occur in the CGHS action, gauge fixed to the
conformal gauge
$(g_{\a\b}=e^{2\rho}\eta_{\a\b})$, through the following relations:
$$Y=\sqrt{2\k}[\r+\k^{-1}e^{-2\phi}-{2\over\k}\int d\phi
e^{-2\phi}\ov h (\phi
)],\eqn\ycdt$$
 $$X=2\sqrt{2\over \k}\int d\phi P(\phi ),\eqn\xcdt$$
where
$$P(\phi )=e^{-2\phi}[(1+\ov h)^2-\k e^{2\phi}(1+h)]^{1\over
2}.\eqn\pee$$
In \ycdt\ and  \xcdt, the functions $h(\phi)$ and $\ov h(\phi)$
parametrize
quantum (measure) corrections that may come in when transforming to
the
translationally invariant measure (see the third paper of  [\sda ]
for
details).  The statement that the quantum theory has to be
independent of the
fiducial metric (set equal to $\eta$ in the above) implies that this
gauge
fixed theory is a CFT. The above solution to this condition was
obtained by
considering only the leading terms of the beta function equations,
but  it can
be shown, in the the cases where  $P$ has no zeroes, \foot{This
implies some
restrictions on the possible quantum corrections, but there is a
large class
which satisfies these conditions.}  that the Liouville-like theory is
an exact
solution to the conformal invariance conditions.

The above considerations mean that there are two classes of quantum
CGHS
models.

a) Those for which $P$ has a zero so that the integration range for
the $X,Y$
fields cannot
be extended over the whole real line.

b) Those for which $P$ has no zeros and are  exact CFTs.

In the case of theories of class (a), one may get  semi-classical
physics which
looks like
spherically symmetric collapse and evaporation in four dimensions by
imposing
reflection boundary conditions where the boundary is timelike [\rst].
There are
however two problems
with this. Firstly, at the semi-classical level the model does not
have a lower
bound to
the ADM energy [\dea ]. This is reflected in the fact that there is a
pulse of
negative energy, the so-called thunderpop[\rst], just before complete
evaporation of the black hole. Secondly, the model (because of the
boundary in
the functional integral) is probably not an exact CFT, and hence it
is unlikely
that it is a true representation of the original quantum CGHS theory.
We will
nevertheless make a few speculative remarks about it in the next
section.

Class (b) gives us an exactly solvable theory of quantum 2d dilaton
gravity.
However it does not give us the picture of four-dimensional
spherically
symmetric black hole evaporation that many authors have been looking
for. The
space-time is conformal to two-dimensional Minkowski space and there
is no
black hole singularity\foot{See the third reference in [\sda]}.
Nevertheless,
as stressed in the work of reference [\shda], it gives us a
theoretical
laboratory in which to study the emergence of semi-classical physics
from the
exact quantum theory.  In [\shda] we attempted to do this starting
from the
Fock space formulation of the theory. Our conclusion was that the
only way to
recover the  semi-classical physics of the theory was to abandon the
standard
procedure of quantization in which the physical state condition was
imposed on
the states, and instead only require that the expectation value in
physical
states of the constraint operators be zero. Although we have not
found any
problem with this approach, it is more satisfying to get the
semi-classical
equations while preserving the superposition principle for the
quantum gravity
wave function.  To this end we discuss the Schr\"odinger wave
functional
approach to the problem with the constraints being implemented
through the WdW
equations for dilaton gravity.

The stress tensor for two-dimensional dilaton gravity can be written
as
$$\eqalign{T_{\pm\pm} &= {1\over 2} (\partial_\pm X \partial_\pm X -
\partial_\pm Y\partial_\pm Y) + \sqrt{\k\over 2} \partial_\pm^2 Y +
{1\over
2}\sum_{i=1}^N \partial_\pm f_i \partial_\pm f_i\cr
T_{+-} &= -\sqrt{\k\over 2} \partial_+\partial_- Y - \lambda^2
e^{\sqrt{2\over
\k}\ (X+Y). }\cr}$$Defining
$$\zeta_+ = \sqrt{2\over \k}\ (X+Y) + \ln {2\over \k}\qquad \zeta_- =
\sqrt{2\over \k}\ (X-Y) $$the components are
$$\eqalign{T_{\pm\pm} &= {\k\over 4}\left[ \partial_\pm \zeta_+
\partial_\pm
\zeta_- +  \partial_\pm^2 (\zeta_+ - \zeta_-)\right] + {1\over
2}\sum_{i=1}^N
\partial_\pm f_i \partial_\pm f_i\cr
T_{+-} &= -{\k\over 4} \partial_+\partial_-  (\zeta_+ - \zeta_-)
-{\k\over2}\lambda^2 e^{\zeta_+ }.\cr}$$The Hamiltonian density is:
$$\eqalign{T_{00}  &= T_{++} + T_{--} + 2 T_{+-}\cr
&= {\k\over 8} \left[\dot\zeta_+ \dot\zeta_-  + \zeta'_+ \zeta'_-  +
2(\zeta''_+  - \zeta''_-) - 8\lambda^2 e^{\zeta_+}\right] + {1\over
4}\sum_{i=1}^N\left[\dot f_i^2 + {f'_i}^2\right] }$$where $\cdot$ and
$'$
denote differentiation with respect to the space-time coordinates
$\tau$ and
$\sigma$ respectively.

Now using the canonical momenta
$$\Pi_\pm = {\k\over 32\pi} \dot \zeta_\mp \qquad \Pi_{f _i}= {1\over
8 \pi}
\dot f_i \eqn\CanPDG$$we obtain
$$T_{00} = {128\pi^2\over \k} \Pi_+ \Pi_- + {\k\over 8}\left[\zeta'_+
\zeta'_-
+
2(\zeta''_+  - \zeta''_-) - 8\lambda^2 e^{\zeta_+}\right] + {1\over
4}\sum_{i=1}^N\left[ {f'_i}^2 + 64\pi^2 \Pi^2_{f_i}\right].\eqn\Tzz$$
We also have the momentum density
$$T_{01} =4\pi\left[\zm' \Pi_-+ \zp'\Pi_+  + 2{\pa\over \pa \s}
\left(\Pi_- -
\Pi_+ \right)+ \sum_{i=1}^N f'_i \Pi_{f_i}\right].\eqn\Tzo$$

We now quantize using $\Pi_u \rightarrow {1\over i} {\delta\over
\delta u}$ and
obtain
the corresponding operators $\hat T_{00}$ and $\hat T_{01}$.
The Wheeler-deWitt equation is thus\foot{ Several authors have
discussed the
WdW equation for dilaton gravity (see reference [\hori] for example),
but none
of them have focused on the problem that we are addressing.}

$$ \hat T_{00} \Psi\left[\zeta_+,\zeta_-,f_i\right] =
0,\eqn\wdwdg$$and we also
have the spatial diffeomorphism constraint
$$ \hat T_{01} \Psi\left[\zeta_+,\zeta_-,f_i\right] =
0.\eqn\wdwdgdiffeo$$
Writing $\Psi$ in the  form $$\Psi = R\, [\zeta_+,\zeta_-,f_i]
\exp{\left\{i
S\,[\zeta_+,\zeta_-,f_i] \right\}}$$ we obtain for the real part of
$\hat
T_{00} \Psi =0$,
$${128\pi^2\over \k} {\d S \over \d\zp} {\d S \over \d\zm}  + \k\
V[\zeta_+,\zeta_-] + V_m[f']  + Q +  16 \pi^2 \sum_{i=1}^N \left({\d
S \over \d
f_i}\right)^2 = 0\eqn\ReTzz$$with
$$V[\zeta_+,\zeta_-] = {1\over 8} \left[\zeta'_+\zeta'_- +
2(\zeta''_+  -
\zeta''_-) - 8\lambda^2 e^{\zeta_+}\right]$$
$$V_m[f']={1\over 4}\sum_{i=1}^N {f'_i}^2 $$
$$Q =- {128\pi^2\over \k} {1\over R} {\d^2 R\over \d\zp \d\zm}
-{16\pi^2\over
R}\sum_{i=1}^N {\d^2 R\over \d f_i^2} .$$
The imaginary part of  $\hat T_{00} \Psi =0$ gives
$$ \sum_{i=1}^N {\d \over \d f_i} \left(R^2 {\d S\over \d f_i}\right)
+ {4\over
\k} \left[  {\d \over \d \zp}\left(R^2 {\d S\over \d \zm}\right) +
{\d \over \d
\zm}\left(R^2 {\d S\over \d \zp}\right)\right] = 0.\eqn\ImTzz$$

For the real and imaginary parts of $\hat T_{01} \Psi = 0$ we have
$$\left(\zm' + 2 {\partial\over\partial \s}\right) {\d R\over \d \zm}
+
\left(\zp' - 2 {\partial\over\partial \s}\right) {\d R\over \d \zp} +
\sum_{i=1}^N f'_i \ {\d R\over \d f_i} = 0,\eqn\ReTzo$$
$$\left(\zm' + 2 {\partial\over\partial \s}\right) {\d S\over \d \zm}
+
\left(\zp' - 2 {\partial\over\partial \s}\right) {\d S\over \d \zp} +
\sum_{i=1}^N f'_i \ {\d S\over \d f_i} = 0,\eqn\ImTzo$$respectively.

The $O(M^{ -1})$ expansion in the  mini-superspace example is
replaced here by
an expansion in $\k^{ -1}$:
$$\eqalign{S[\zeta_+,\zeta_-,f_i] &=\k S_{-1}[\zeta_+,\zeta_-,f_i] +
S_0[\zeta_+,\zeta_-,f_i]  + {1\over \k}S_1[\zeta_+,\zeta_-,f_i] +
\ldots\cr
R[\zeta_+,\zeta_-,f_i] &= R_0[\zeta_+,\zeta_-,f_i]  + {1\over
\k}R_1[\zeta_+,\zeta_-,f_i] + \ldots.\cr}$$
One finds
$$\eqalignno{{\cal O}(\k^2): \qquad &0={\delta S_{-1}\over \delta
f_i}&\eqnalign{\DGONN} \cr\cr
{\cal O}(\k ): \qquad &0=128 \pi^2 {\delta S_{-1}\over \delta
\zeta_+} {\delta
S_{-1}\over \delta \zeta_-}  + {1\over 8} \left(\zeta'_+ \zeta'_- + 2
(\zeta''_+ - \zeta''_-) - 8\lambda^2
e^{\zeta_+}\right).&\eqnalign{\DGON}
\cr}$$As in the mini-superspace example one finds that largest
component of $S$
is independent of the matter fields, and the metric is determined by
the vacuum
Einstein equation (equation \DGON).

For the ${\cal O}(1)$ equations we find
$$\eqalign{&{\rm Real:}\quad V_m[f'] + 16\pi^2
\sum_{i=1}^N\left({\delta
S_0\over \delta f_i}\right)^2 + 128 \pi^2 \left[{\delta S_{-1}\over
\delta
\zeta_+}{\delta S_{0}\over \delta \zeta_-} +
{\delta S_{-1}\over \delta \zeta_-}{\delta S_{0}\over \delta
\zeta_+}\right] -
16\pi^2 \sum_{i=1}^N{1\over R_0} {\delta^2 R_0\over \delta f_i^2} =
0\cr
&{\rm Imaginary:}\quad \sum_{i=1}^N {\delta \over \delta f_i}
\left[R_0^2
{\delta S_0\over \delta f_i}\right] + 4 \left[{\delta\over \delta
\zeta_+}
\left(R_0^2 \,{\delta S_{-1}\over \delta \zeta_-}\right) +
{\delta\over \delta
\zeta_-} \left(R_0^2 \,{\delta S_{-1}\over \delta \zeta_+}\right)
\right] =
0\cr}$$
One now defines a (local) time by $${\delta\over \delta T}
=128\pi^2\left\{
{\delta S_{-1}\over \delta \zeta_+}{\delta \over \delta \zeta_-} +
{\delta S_{-1}\over \delta \zeta_-}{\delta \over \delta
\zeta_+}\right\}.\eqn\TDef$$The real equation becomes
$${\delta S_0\over \delta T}= -V_m[f'] - 16\pi^2
\sum_{i=1}^N\left({\delta
S_0\over \delta f_i}\right)^2  +  16\pi^2 \sum_{i=1}^N{1\over R_0}
{\delta^2
R_0\over \delta f_i^2} .$$

Now given a function $g$ (independent of the $\{f_i\}$) satisfying
$${1\over g} {\delta g\over \delta T} = 128\pi^2{\delta^2 S_{-1}\over
\delta
\zeta_+ \delta \zeta_-},$$and defining $r_0$:$$r_0 = R_0 g,$$ the
imaginary
equation becomes the continuity equation
$$ {\delta r_0\over \delta T}=-{16\pi^2\over r_0} \sum_{i=1}^N
{\d\over \d f_i}
\left( r_0^2 {\d S_0\over \d f_i}\right).$$Writing $$\chi = r_0 e^{i
S_0}$$we
have the Schr\"odinger  equation $$i {\delta \chi\over \delta T} =
{\cal H}_m
\chi$$with
$$ {\cal H}_m = -16\pi^2\sum_{i=1}^N{\delta^2\over \delta f_i^2} +
V_m[f'].$$

Thus again we have the Schr\"odinger equation for matter, in a
background
determined by the vacuum Einstein equation.

In order to see whether, as in the mini-superspace example, one can
at least
give a heuristic argument to justify the usual  semi-classical
equations we
proceed as follows.
Define
$$\Psi[\zp,\zm,f_i] = F[\zp,\zm]\ \chi[\zp,\zm,f_i]\ \exp\left\{i\k
S_{-1}[\zp,\zm]\right\}.$$
Substitution in the WdW equation \wdwdg\ gives (using equation \TDef)

\vbox{
$$\left(\hat T^m_{00} + {128\k \pi^2} {\d S_{-1}\over \d \zp}{\d
S_{-1}\over \d
\zm} + \k V_G\right)F \chi -  i  {\d \over \d T} (F \chi)$$
$$- \ {128\pi^2\over \k} {\d^2\over \d \zp \d\zm} (F \chi) - 128 i
\pi^2
F\chi \   {\d^2 S_{-1}\over \d \zp \d\zm} = 0\eqn\WdWDone$$
}with$$\hat T^m_{00} \equiv \sum_{i=1}^N\left\{-16\pi^2 {\d^2\over \d
f_i^2} +
{1\over 4}  f'^2_i\right\}.$$

We now take $S_{-1}$ to be a $f_i$ independent solution of  the
semi-classical
equation
$${128\k \pi^2} {\d S_{-1}\over \d \zp}{\d S_{-1}\over \d \zm} + \k
V_G =
-\vev{\hat T^m_{00}}\eqn\SCDil$$
and equation \WdWDone\ becomes

\vbox{
$$\left(\hat T^m_{00} - \vev{\hat T^m_{00}}\right) \chi - i {\d
\chi\over\d
T}-{128i \pi^2}\left( {\d^2 S_{-1}\over \d \zp\d\zm} - {i  \over \k}
{1\over
F}{\d^2 F\over \d \zp\d\zm} + {1\over 128 \pi^2}{1\over F} {\d F\over
\d T}
\right)\chi $$
$$- \ {128\pi^2\over\k}\left({1\over F} {\d F\over \d \zp}{\d
\chi\over \d \zm}
+
{1\over F} {\d F\over \d \zm}{\d \chi\over \d \zp}  - {\d^2 \chi\over
\d
\zp\d\zm}\right) = 0.\eqn\WdWDtwo$$
}In this case the functional $F$ will be defined by requiring that
the
coefficient of $\chi$ in the third term be zero:
$${\d^2 S_{-1}\over \d \zp\d\zm} - {i  \over \k} {1\over F}{\d^2
F\over \d
\zp\d\zm} + {1\over 128 \pi^2}{1\over F} {\d F\over \d T}  =0.$$

We now want to determine under what conditions equation \WdWDtwo\
reduces to
the Schr\"odinger-like  equation
$$\left(\hat T^m_{00} - \vev{\hat T^m_{00}}\right) \chi =i{\d
\chi\over\d
T}.$$In particular, note that the extra terms $$ -\
{128\pi^2\over\k}\left({1\over F} {\d F\over \d \zp}{\d \chi\over \d
\zm} +
{1\over F} {\d F\over \d \zm}{\d \chi\over \d \zp} - {\d^2 \chi\over
\d
\zp\d\zm} \right) $$will become small as $\k$ ($N$) gets large.
However in the
same limit our semi-classical equation \SCDil\ will approach the
vacuum
Einstein equation and the usual formulation cannot be recovered.

One might ask whether the required equations can be obtained if
$\vev{\hat
T^m} $ scales with $\k$ so that the left hand side of \SCDil\ is of
the same
order in the expansion as the right hand side. However then we have
no
justification for separating \WdWDtwo\ as the $O(\k^0)  $ terms. In
other words
if the large $\k$ approximation is to be used it must be done
systematically
and that just leads us to the previous results, i.e. the
Schr\"odinger equation
in the vacuum background. As far as we can see there is no
alternative argument
(as in the mini-superspace case) either.

\chapter{Other Possibilities}
\section{ Reduced phase space quantization (light cone gauge)}
An alternative to Dirac quantization is to first solve the classical
constraints and then quantize the physical dynamical variables.  In
this case
one has an intrinsic (local) time in terms of which the wave
function(al) of
the physical variables satisfies a local functional  Schr\"odinger
equation. It
may then be thought that this gives us a definition of time in terms
of which
the $S $(or $\$$) matrix can be defined. However, as discussed in
some detail
by Kucha\v{r} [\kk ], this method does not give a resolution to the
problem of
time: It is beset with the same difficulties as the Dirac
quantization method,
in addition to suffering from problems like the multiple choice
question
stemming from the non-uniqueness in the choice of the time variable.
Also the
Hamiltonian that is obtained from solving the classical constraints
is  in
general non-local\foot{Though in the case of dilaton gravity it is
possible to
have a local Hamiltonian.}. More importantly, we cannot find any way
of
reproducing the semi-classical physics of black hole collapse and
Hawking
radiation starting from this version of the exact quantum theory.

It should also be stressed at this point that the mere existence of a
Hermitian
reduced (local) Hamiltonian is not at all a solution of the problem
posed by
Hawking. Firstly, since the theory does not admit the usual
semi-classical
picture it is not clear how the problem can even be posed. Secondly,
although a
Hermitian Hamiltonian will lead to unitary evolution, the point is
that
according to the semi-classical picture part of the state has gone
into the
black hole and cannot be reconstructed by an asymptotic observer
outside the
hole. This is the origin of information loss according to Hawking
[\hawk ].
Any refutation of Hawking's claim should at least first reproduce
this
semi-classical picture in the appropriate regime.

\section{Boundary Hamiltonian}
 It is well known  that the classical  Hamiltonian for a space-time
that is
asymptotically flat (such as that of a black hole) has a boundary
contribution.
Since the bulk Hamiltonian is weakly zero in a generally covariant
theory, the
total energy (ADM mass) of such configurations is given by the value
of this
boundary term. It has been shown by Regge and Teitelboim [\rt] that
this term
is necessary in order to cancel a boundary term that arises  in the
derivation
of Hamilton's equations for the system.

In a local quantum field theory however such a boundary Hamiltonian
cannot play
any
role whatsoever. {\em Indeed microcausality requires that all local
fields will
commute with such a boundary Hamiltonian} and {\em it is irrelevant
to the
derivation of the Heisenberg equations.  The time evolution of the
quantum
field theory  does not depend on such a boundary Hamiltonian.}  To
put it
another way one must define the theory with an infra-red cut-off (or
by
smearing with test functions which fall off rapidly at spatial
infinity). Thus
there will be no boundary  contributions to the energy. One should
not expect
the S-matrix (or \$ matrix) to be defined by some quantum analogue of
the ADM
energy.  The latter seems to have meaning only within the
(semi-)classical
context.

\chapter{de Broglie-Bohm Interpretation}

When one applies quantum mechanics to the universe as a whole, the
usual
pragmatic interpretation, which separates the world into classical
observing
system and quantum
systems, is clearly untenable. This has indeed been a serious
conceptual
barrier (quite
distinct from the technical problems of quantum gravity) to
understanding
quantum gravity. One possibility is to adopt the so called Everett
(or many
worlds) interpretation. However, it is not at all clear that this
interpretation provides us with an explanation of why it is that
experimental
results have definite values.\foot{Recent work by Gell-mann and
Hartle [\gh],
as well as earlier work cited in their work may clarify these issues,
but we do
not understand these works sufficiently well to apply them to our
problem.}
The other possiblity (recommended to cosmologists and by implication
practitioners of quantum gravity by John Bell [\jb ]) is the de
Broglie-Bohm
(deBB) interpretation\refmark{\holland}.

We have argued in this paper (and in [\shda]) that none of the usual
arguments
for deriving the semi-classical picture from an exact formulation of
quantum
gravity are valid. As we've seen the problem is essentially the
problem of time
that has been discussed extensively in the literature on quantum
gravity [\kk].
The deBB picture gives a natural way of getting parametric time in
quantum
gravity [\vink, \holland], and it is the only way that we have found
where one
can establish the validity of the physical picture of the
semi-classical
calculations.  After reviewing very briefly the deBB formulation of
one
particle quantum mechanics, we will discuss the deBB interpretation
of quantum
mini-superspace.   Next, we will work out the dilaton gravity case
and discuss
under what conditions the semi-classical calculations might be valid.

Substituting the  form $\psi(x,t) = R(x,t) \exp{i S(x,t)}$ into the
Schr\"odinger  equation, with $R$ and $S$ real functions, one has
from the real
and imaginary equations:
$$\eqalignno{{\partial S\over\partial t} + {(\partial_x S)^2\over 2
m} + V + Q
&=0&\eqnalign{\HJeq}\cr
{\partial R^2\over \partial t} + \partial_x \left(R^2 \partial_x
S\right)
&=0&\eqnalign{\Prob}\cr}$$with $$Q= -{\hbar^2\over 2m}{\partial_x^2
R\over
R}\qquad R=|\psi|^2.\eqn\QPot$$

In the deBB interpretation one now postulates that the particle moves
on a
trajectory $X(t)$ with momentum given by the Hamilton-Jacobi formula
$$P_x = m\dot{X}=\partial_x S|_{x=X}.$$
When this is used in equation \HJeq, it is in the form of the
classical
Hamilton-Jacobi equation, except that there is an additional
``quantum
potential" term $Q$.  By differentiating \HJeq\ one obtains the
classical
equations of motion corrected by quantum terms coming from the $Q$
term. Once
the initial wave function and the initial position of the particle
are given,
the theory predicts the wave function and particle position at any
future time.
The wave function itself plays a dual role. Firstly, it affects the
particle
motion through the $Q$ term in the equations of motion. Secondly, it
gives the
distribution of possible initial values of the particle position, and
hence as
a result of the continuity equation \QPot, also the distribution of
positions
at any future time.  The interpretation can be shown  to be in
agreement  with
all experimental tests of quantum mechanics.  Its great merit is that
it avoids
the ambiguities and paradoxes associated with the pragmatic
Copenhagen
interpretation.  This includes in particular the dividing line
between the
classical and quantum realms, and the mysterious (non-unitary)
collapse of the
wavefunction. In effect it is an observer independent realist
interpretation,
and it seems to us that the deBB formulation, or something along
those lines,
is essential for the discussion  not only of  quantum cosmology, but
also of
the physics of black holes.

Let us now describe the deBB interpretation in the case of
mini-superspace
\foot{These equations have already been derived by Vink [\vink].
However his
interpretation seems to be rather different since he seems to use
them as  a
step towards the derivation of the usual semi-classical equations
\SCone\ and
\Sch, which in our opinion are untenable except under very special
conditions
as outlined in section 2, and for this purpose we do not think it is
necessary
to invoke the deBB interpretation.}.
We rewrite  equation \RePart\  as
$$-{1\over 2 a M^2} (\partial_a S)^2 + {1\over 2 a^3} (\partial_\phi
S)^2 +
a^3\left[M^2 V_G + V_m + Q\right]  = 0 \eqn\dBRePart$$
with $$Q =  { \hbar^2\over 2 a^6 }{1\over R}\left[{1\over M^2}
(a\partial_a)^2
R - \partial_\phi^2 R\right].\eqn\dBQP$$
Using the canonical momenta defined by equation \CanP\ we define  the
trajectories for the scale factor $A(t)$ and the matter field $
\Phi(t)$ by
$${d A(t)\over d t} = {1\over A(t) M^2} \partial_a S \,{\big
|}_{a=A\atop \phi
= \Phi}\qquad \qquad{d \Phi(t)\over d t} = {1\over A(t)^3}
\partial_\phi S
\,{\big |}_{a=A\atop \phi = \Phi}.\eqn\QTraj$$
It should be noted that the equation for the classical  trajectory of
the scale
factor in (2.15) is of exactly the same form as the above equation,
except that
there S was just the solution to the classical Hamilton-Jacobi
equation for $a$
in the absence of matter. Thus time was defined only with respect to
some
vacuum configuration. On the other hand in the deBB interpretation
parametric
time arises in the full quatum theory precisely because choosing a
particular
trajectory amounts to a spontaneous breakdown of general covariance.
It is only
the full ensemble of trajectories described by the wave fuctional
$\Psi  $ that
satisfies general covariance.

Substituting equations \QTraj\ into equation \dBRePart\ yields
$$-M^2 \left({A \dot A^2\over 2} - A^3 V_G\right) + A^3\left({\dot
\Phi^2\over
2} + V_m + Q\right) = 0.$$ As $Q\rightarrow 0$ we obtain the
classical equation
\eqf.

Differentiating \QTraj\ with respect to $t$, and differentiating
\dBRePart\
with respect to $a$, one obtains, after some manipulation:
$$M^2\left\{\partial_t (A \dot A) - {\dot A^2\over 2} - \partial_a
\left(A^3
V_G\right) \right\} + 3 A^2\left({\dot\Phi^2\over 2} - V_m\right)
=\partial_a(A^3 Q).$$
Similarly one gets
$${d\over d t}(A^3 \dot \Phi) + A^3 \partial_\phi V_m = - A^3
\partial_\phi
Q.$$In the limit $Q\rightarrow 0$ these two equations become the
classical
equations of motion \eqs\ and \eqt\ respectively.

The classical equations are obtained in the regime where $Q$ is
negligible. It
should be noted that although this term is explicitly of $O(\hbar^2)$
this is
not sufficient reason to neglect it. Whether one can neglect it or
not is a
delicate question depending on the form of the wave functional. Even
in quantum
mechanics there are states which have no classical limit, such as
stationary
states, for which this term exactly cancels the classical potential
term. In
quantum cosmology an example of this is the Hartle-Hawking wave
function [\hh]
which is real and thus will not allow a classsical limit. It also
will not have
an evolving geometry or matter fields according to the deBB
interpretation; it
is the analogue of a stationary state and is a superposition of
expanding and
contracting states.  The Vilenkin wave function [\vil] on the  other
hand
gives an expanding quantum trajectory with a well defined classical
limit.

The deBB interpretation is extended to field theory in a straight
forward
manner by replacing the trajectory of the particle by trajectories
for the
field variables.  In 2-d dilaton gravity the analogs of equations
\HJeq\ and
\Prob\ are the WdW equations given by \ReTzz\ and \ImTzz.  In
addition we also
have the constraint equations \ReTzo\ and \ImTzo.  The trajectories
are defined
using equation \CanPDG:
$$\eqalign{\dot Z_+(\s,\tau)&\equiv{\pa Z_+(\s,\tau) \over \pa
\tau}= {32 \pi
\over \k} {\d S \over \d \zm}\biggr{|}_{\z_\pm =Z_\pm\atop
f_i=F_i}\cr\cr
\dot Z_-(\s,\tau)&\equiv{\pa  Z_-(\s,\tau) \over \pa \tau}= {32 \pi
\over \k}
{\d S \over \d \zp}\biggr{|}_{\z_\pm =Z_\pm\atop f_i=F_i}\cr\cr
\dot F_i(\s,\tau) &\equiv{\pa  F_i (\s,\tau)\over \pa  \tau}= 8\pi
{\d S \over
\d f_i}\biggr{|}_{\z_\pm =Z_\pm\atop
f_i=F_i}.\cr\cr}\eqn\QTrajDG$${\em These
equations implicitly define a time parameter $\tau$}.
As before we get dynamics only for complex wave functionals ($S\ne
0$).

We functionally differentiate (the spatial integral of) \ReTzz\ with
repect to
$\zp(\s)$, and the second equation in  \QTrajDG\ with respect to time
to
obtain$$  {\k\over 8}\left(\ddot{Z}_-(\s)   - Z''_-(\s) \right) -
\l^2 \k
e^{Z_+(\s)} = -{\d\over \d\zp(\s)}\int Q(\s') d\s'\biggr{|}_{\z_\pm
=Z_\pm\atop
f_i=F_i}.\eqn\dBBa$$
In a similar manner we obtain the quantum extensions of the other
classical
equations of motion:
$${\k\over 8}\left(\ddot{Z}_+   - Z''_+ \right)  = -{\d\over \d\zm}
\int
d\s'Q(\s')\biggr{|}_{\z_\pm =Z_\pm\atop f_i=F_i} \eqn\dBBb$$and
$${1\over 2}\left(\ddot{F_i}   - F''_i \right)  = -{\d\over \d f_i}
\int
d\s'Q(\s')\biggr{|}_{\z_\pm =Z_\pm\atop f_i=F_i}.\eqn\dBBc $$

In conjunction with the above equations of motion one also has, after
substituting \QTrajDG\ into \ReTzz\ and \ImTzo,  the de Broglie-Bohm
versions
of the constraints
$${\k\over 8} \left( \dot Z_+ \dot Z_- + Z'_+ Z'_- + 2\left(Z''_+ -
Z''_-\right) - 8 \l^2 e^{Z_+}\right) + {1\over 4}
\sum_{i=1}^N\left({F'_i}^2 +
{\dot F_i}^2\right)  = -Q\eqn\dBBCone$$
$${\k\over 4} \left[ \left(Z'_- + 2 {\partial\over \partial
\s}\right) \dot Z_+
+ \left(Z'_+ - 2 {\partial\over \partial \s}\right) \dot Z_- \right]
+
\sum_{i=1}^N F'_i \dot F_i=0.\eqn\dBBCtwo$$

These are to be compared with the classical equations of motion
derived from
the action \action$$\eqalign{  \ddot \zeta_- - \zeta''_-  =&  8
\lambda^2
e^{\zeta_+}\cr
\ddot \zeta_+ - \zeta''_+ =& 0 \cr \ddot f_i - f''_i =& 0,
\cr}\eqn\DGClassEOM$$ and the constraint equations coming from \Tzz\
and \Tzo.

When $Q\rightarrow 0$, equations \dBBa\ - \dBBCtwo\ reduce to the
semi-classical equations discussed in [\sda, \bc, \rst] and many
other
works\foot{We call these semi-classical rather than classical
equations because
the definitions of the fields incorporate $O(\k\hbar) $ conformal
anomaly
corrections.  It should also be noted that if we start from the WdW
equation of
the original classical theory, then ${\cal O}(\hbar)$ corrections,
which are
responsible for Hawking radiation, are missing.}. The question as to
what
extent the semi-classical equations are valid then becomes one of
deciding when
and in what regimes the quantum potential term becomes negligible. As
pointed
out earlier in general this cannot be done for any wave functional
$\Psi$. In
particular since the Wheeler-deWitt equation is real it is always
possible to
find a real solution and, as in the stationary state of quantum
mechanics or
the Hartle-Hawking wave function discussed above, there will not be
any time
evolution. In order to have a dynamic semi-classical picture, one
needs a
complex wave functional.

Given  a complex wave functional one may then ask under what
conditions $Q$ is
negligible. Unfortunately we have not yet been able to solve the WdW
equation
for the
theory in order to decide this issue.\foot{The solutions that exist
in the
literature [\hori] do not include the conformal anomaly corrections
and hence
it is not clear how to obtain the comparison to the calculations of
Hawking
radiation. }  We can however speculate that, since the theories with
no
boundary in field space (see section 3 and [\rst] and [\shda]) have
effective
actions which are the same as the (semi-)classical actions, the wave
functional
is a pure phase so that $R=const.$ and $Q$ is zero. On the other hand
it is
likely that in theories with boundary, as in that described by
[\rst], $Q$ is
in fact significant, especially when one approaches the boundary of
field space
(which  in the semi-classical analysis becomes space-like and hence
gives rise
to information loss in this analysis). It should also be pointed out
that $Q$
incorporates all the EPR type non-localities of quantum mechanics and
may well
account for the mechanism by which information is extracted from
behind the
horizon of the semi-classical picture.  It may also completely
vitiate the
semi-classical picture in that the black hole itself may be washed
out. These
are questions that are presently under investigation.

\chapter{Conclusions}
Generally covariant quantum field theories are topological theories
in so far
as all correlation functions are independent of space-time geometry
and the
only physical operators are those which commute with the constraints.
The
theory will thus  only contain topological information [\wit]. The
problem of
time in quantum gravity is just a manifestation of this fact. It
appears then
that only the spontaneous breaking of general covariance at the
quantum level
can restore a geometrical background within which physical processes
can be
discussed.  One of the conclusions of this work is that the work on
semi-classical gravity (quantum field theory in curved space-time)
has to be
interpreted in this way.

In the particular case of black hole formation and evaporation it has
been
argued that these questions are irrelevant. We find though that this
is not the
case. First we showed, by a detailed discussion of both
mini-superspace and
dilaton gravity, that there is no systematic derivation of quantum
field theory
in a background that is determined by the Einstein equation with a
source term
given by the expectation value of the matter stress tensor. What one
gets from
the expansion in inverse powers of the Planck mass or the large $N$
expansion
is quantum field theory for the matter sector in a background that is
a
solution of the vacuum Einstein equations [\banks].  Thus for
instance, the
usual applications of QFT in flat space particle physics for energy
densities
that are small compared to the Planck density are justified. One may
argue in
the mini-superspace case that, while there is no systematic
derivation of the
so-called semi-classical equations, it is still possible to show that
in the
late universe  the functional Schr\"odinger equation is consistent
with that
which arises from the Wheeler-deWitt equation, once a classical
cosmological
solution (determined by the expectation value of the matter solution)
is used.
This argument perhaps justifies quantum field theory calculations in
cosmological backgrounds. However, for 2d  dilaton gravity, the only
known
field theoretic example that exhibits (at the semi-classical level)
the
formation and evaporation of black holes, we cannot find such an
argument.

We are then left with the problem of justifying the semi-classical
equations of
dilaton gravity that many authors \foot{See [\reviews] for reviews.}
have used
to explore the problems associated with Hawking radiation. The
equations in
question involve space-time differential equations for c-number
fields. In an
earlier paper by one of us [\shda], it was argued that if  these  are
to be
interpreted as expectation values of the corresponding equations for
quantized
fields then one is forced to drop the constraint equation (WdW or
BRST) as an
equation on the states. In other words, the semi-classical equations
cannot
emerge as expectation values, in generally covariant physical states,
of  the
corresponding operator equations. One way out suggested in that paper
was to
abandon the superposition principle at the level of the wave function
of the
universe, and to impose the constraint by defining physical states as
those in
which the constraint is satisfied as an expectation value.   Although
we do not
see any logical flaw in this, in this paper we have examined an
alternative
which enables us to obtain the required semi-classical physics from
the WdW equation. This involves  the de Broglie-Bohm interpretation
of quantum
mechanics.

It seems to us that any discussion of quantum gravity must go beyond
the
pragmatic interpretation that divides the world into classical
apparatus and
quantum system. This is certainly the case for quantum cosmology, but
the
arguments presented
in this paper indicate that this is so even for the black hole
problem, at
least if we start from generally covariant physical states. As far as
we can
see, the only way to get a picture of the evolution of a black hole
is to pick
a quantum trajectory (in the sense of deBB) from the ensemble of
trajectories
that may be described by the Schr\"odinger  wave functional. This
constitutes a
spontaneous breakdown of general covariance.  The resulting equations
of motion
and constraint equations are precisely the semi-classical equations
discussed
in the literature except that there are correction terms. Given a
solution of
the WdW equation, these additional terms may be evaluated, and to the
extent
that they are small, one may say that the semi-classical
approximation is
valid. We have argued that for the dilaton gravity theories with no
field space
boundary, these correction terms are (probably) absent. However in
this case
there is no black hole singularity and no information loss problem.
\foot{Information is lost only in the trivial sense that left movers
carry
information from the right end of space to the left end and do not
communicate
with the right movers. This is a situation that one may obtain in
ordinary flat
space quantum mechanics and does not imply non-unitary evolution. }
On the
other hand, in the theories with boundary, such as that of RST
[\rst], it is
not at all clear that one should ignore these correction terms. We
are
presently engaged in finding solutions to the WdW equation for such
models so
that these issues may be further clarified.

\chapter{Acknowledgements}
This work is partially supported by the Department of Energy contract
No.
DE-FG02-91-ER-40672.

\vfill\eject

\refout

\end